\documentclass[12pt]{article} 
\usepackage{amsmath,amssymb}
\usepackage[english]{babel}
\title{Feynman's derivation of Maxwell equations and extra dimensions}
\author{ Z.~K.~Silagadze 
\vspace*{3mm} \\
Budker Institute of Nuclear Physics,  630 090,
Novosibirsk, Russia }
\date{}

\begin{document}
\large
\maketitle

\begin{abstract}
It is shown that Feynman's derivation of Maxwell equations admits 
a generalization to the case of extra spatial dimensions. The generalization
is unique and is only possible in seven dimensional space.
\end{abstract}

%\newpage

\section{Introduction}
Some times ago Dyson published a paper \cite{1} about unusual proof of
homogeneous Maxwell equations that Feynman had shown him in 1948. Feynman 
himself
``refused to publish it then because he claimed it was only a joke'' 
\cite{2}. Subsequent reaction of the scientific community, in both immediate
[3--11] and long terms [12--15], shows that ``the joke'' was appreciated, and 
scrutinized with great seriousness.

According to Dyson \cite{1}, Feynman's ``purpose was to discover a new 
theory, not to reinvent the old one'' and he did not published the 
derivation because ``his proof of the Maxwell equations was a demonstration 
that his program had failed.''

In this note I intend to show that one could really end with a unique
new theory if we ask a simple question about multi-dimensional generalization
of Feynman's arguments. This question is quite natural in view of present
interest in extra dimensions. Of course the Lorentz covariant version of
Maxwell equations admits an obvious n-dimensional generalization. In fact
Tanimura had already given \cite{8} an n-dimensional generalization of
Feynman's proof along these lines. But such generalizations are against
the spirit of Feynman's original derivation because they involve additional
assumptions which make the proof relativistic and orthodoxly sterile.  

Relativity is a sacred caw of modern physics. But it may happen
that the Lorentz invariance of our low-energy world is violated at higher
energies \cite{16} where presumably extra dimensions open. The condensed
matter analogy shows \cite{17} that both the special and general relativity
might be just emergent phenomena valid only in the low energy limit. 
Therefore it is by no means obvious that Lorentz covariance is the proper
guide when going to extra dimensions. So we will rely on the original
non-relativistic arguments and try to modify them as little as possible.
   
\section{Feynman's derivation of Maxwell equations}
In fact Feynman's derivation is quite heretical as it mixes classical and 
quantum concepts. But the proof has a great virtue of being impressive and
unexpected. So we close eyes to this heresy. The starting points are 
Newton's second law
\begin{equation}
m\dot {\vec{v}}=\vec{F}(x,v,t)
\label{eq1} \end{equation}
\noindent and commutation relations between position and velocity
\begin{equation}
[x_i,x_j]=0, \; \; \; [x_i,v_j]=i\frac{\hbar}{m}\delta_{ij}.
\label{eq2} \end{equation}
These commutation relations imply that for any functions $f(x,v,t)$ and
$g(x,t)$ one has
\begin{equation}
[x_i,f(x,v,t)]=i\frac{\hbar}{m}~\frac{\partial f}{\partial v_i}, \; \;
[v_i,g(x,t)]=-i\frac{\hbar}{m}~\frac{\partial g}{\partial x_i}.
\label{eq3} \end{equation}
Differentiating the second equation in (2) with respect to time and using 
(1) we get
$$[v_i,v_j]+\frac{1}{m}[x_i,F_j]=0$$
or
\begin{equation}
\frac{\partial F_j}{\partial v_i}=\frac{im^2}{\hbar}[v_i,v_j].
\label{eq4} \end{equation}
R.h.s. of this equation does not depend on velocity. Indeed, equations 
(2),(3) and the Jacobi identity imply
$$\frac{\partial}{\partial v_k}[v_i,v_j]\sim [x_k,[v_i,v_j]]=
-[v_i,[v_j,x_k]]-[v_j,[x_k,v_i]]=0.$$
Let us define the field $B(x,t)$ by equation (summation over repeated
indexes is assumed as usual)
$$\frac{im^2}{\hbar}[v_i,v_j]=-\epsilon_{ijk} B_k$$
which implies
\begin{equation}
B_i=-\frac{im^2}{2\hbar}\epsilon_{ijk} [v_j,v_k].
\label{eq5} \end{equation}
Then equation (4) can be integrated with the result
\begin{equation}
\vec{F}(x,v,t)=\vec{E}(x,t)+<\vec{v}\times\vec{B}(x,t)>.
\label{eq6} \end{equation}
Here $<\vec{v}\times\vec{B}>=\frac{1}{2}(\vec{v}\times\vec{B}-
\vec{B}\times\vec{v})$ and reflects our use of symmetric Weyl-ordering 
to resolve operator-product ordering ambiguities.

So we get Lorentz force law if one can identify $E$ and $B$ with electric
and magnetic field respectively. To do such identification, however, one 
should show that $E$ and $B$ obey maxwell equations. We have no problems
with $\mathrm{div}\vec{B}$:
$$\mathrm{div}\vec{B}\sim[v_i,B_i]\sim \epsilon_{ijk}[v_i,[v_j,v_k]]=0.$$
The last step is due to the Jacobi identity
$$\epsilon_{ijk}[v_i,[v_j,v_k]]=\frac{1}{3}\epsilon_{ijk} \left (
[v_i,[v_j,v_k]]+[v_j,[v_k,v_i]]+[v_k,[v_i,v_j]] \right ) =0.$$
To prove the second Maxwell equation we rewrite (5) as 
$$\vec{B}=-\frac{im^2}{\hbar}~\vec{v}\times \vec{v}$$
and calculate the total time-derivative 
$$\frac{d\vec{B}}{dt}=\frac{\partial\vec{B}}{\partial t}+
<\vec{v}\cdot \nabla \vec{B}>=-\frac{im^2}{\hbar}~(\dot{\vec{v}}\times \vec{v}
+\vec{v}\times\dot{\vec{v}})=-\frac{im}{\hbar}~(\vec{F}\times \vec{v}+
\vec{v}\times\vec{F})=$$
$$=-\frac{im}{\hbar}~\left ( \vec{E}\times \vec{v}+\vec{v}\times\vec{E}+
<\vec{v}\times\vec{B}>\times~ \vec{v}+\vec{v}~\times<\vec{v}\times\vec{B}>
\right ).$$
Here again Weyl-ordering is assumed when the time derivative is calculated:
$$<\vec{v}\cdot \nabla \vec{B}>_i=\frac{1}{2}\left ( v_j\frac{\partial B_i}
{\partial x_j}+\frac{\partial B_i}{\partial x_j} v_j \right ).$$

\noindent But the second equation in (3) implies
$$-\frac{im}{\hbar}~\left (\vec{E}\times \vec{v}+\vec{v}\times\vec{E}
\right )=-{\mathrm rot}\vec{E}.$$
As to the remaining terms, we have
$$[<\vec{v}\times\vec{B}>\times ~\vec{v}+\vec{v}~\times<\vec{v}\times\vec{B}>]
_i=\epsilon_{ijk}[v_j,<\vec{v}\times\vec{B}>_k]=$$ 
$$=\frac{1}{2}\epsilon_{ijk}\epsilon_{kmn}[v_j,v_mB_n+B_nv_m]=
\frac{1}{2}[v_j,v_iB_j+B_jv_i-v_jB_i-B_iv_j]=$$
$$=\frac{1}{2}[v_j,v_i]B_j+\frac{1}{2}B_j[v_j,v_i]-\frac{1}{2}v_j[v_j,B_i]-
\frac{1}{2}[v_j,B_i]v_j.$$
The first two terms give zero contribution:
$$[v_j,v_i]B_j+B_j[v_j,v_i]\sim \epsilon_{ijk}(B_kB_j+B_jB_k)=0.$$
The last two terms, on the other hand, give
$$-\frac{1}{2}v_j[v_j,B_i]-\frac{1}{2}[v_j,B_i]v_j=
i\frac{\hbar}{m}~\frac{1}{2}
\left (v_j\frac{\partial B_i}{\partial x_j}+\frac{\partial B_i}{\partial x_j}
v_j \right )=i\frac{\hbar}{m}<\vec{v}\cdot\nabla \vec{B}>_i \; .$$ 
Therefore
$$\frac{\partial\vec{B}}{\partial t}+<\vec{v}\cdot \nabla \vec{B}>=
-{\mathrm rot}\vec{E}+<\vec{v}\cdot \nabla \vec{B}>$$
and we recover the Faraday law.

As we see, starting from Newton's non-relativistic law of motion and
commutation relations between position and velocity we really managed to 
obtain Maxwell equations -- ``a thing which baffles everybody including
Feynman, because it ought not be possible'' \cite{2}. Concepts are clear,
mathematics simple, but you do not believe the result, do you? You do not 
believe that it is possible from non-relativistic presumptions to obtain
truly relativistic equations. And you are right. In fact we have derived
only half of Maxwell equations -- the Bianchi set:  
\begin{equation}
\mathrm{div}\vec{B}=0,\;\;\; \frac{\partial\vec{B}}{\partial t}+
{\mathrm rot}\vec{E}=0.
\label{eq7} \end{equation}
These equations are indeed compatible with Galilean invariance \cite{5}.
The whole set of Maxwell equations, however, cannot coexist peacefully
with the Galilean transformations due to presence of the ``displacement
current'' term  \cite{5}. But here we are not interested in such kind of
subtleties. Instead our prime interest lies in multi-dimensional 
generalization of Feynman's arguments. To succeed in such an enterprise
we need some notion of vector product in multi-dimensional space. This is
particularly clear from the final form (7) of Maxwell equations.
 
\section{Vector product in multi-dimensional space}
Let us consider n-dimensional vector space $\mathbb{R}^n$ over the real
numbers. What properties we want the multi-dimensional vector product in 
$\mathbb{R}^n$ to satisfy? Intuitively it is reasonable to demand 
$$(\vec{A}\times\vec{B})\cdot \vec{A}=(\vec{A}\times\vec{B})\cdot \vec{B}=
0$$ and $\vec{A}\times\vec{A}=0$. But then $$0=(\vec{A}+\vec{B})\times
(\vec{A}+\vec{B})=\vec{A}\times\vec{B}+\vec{B}\times\vec{A}$$ shows that
the vector product is anti-commutative. By the same trick one can prove
that $(\vec{A}\times\vec{B})\cdot\vec{C}$ is alternating in 
$\vec{A},\vec{B},\vec{C}$. For example $$0=((\vec{A}+\vec{C})\times\vec{B})
\cdot (\vec{A}+\vec{C})=(\vec{C}\times\vec{B})\cdot\vec{A}+
(\vec{A}\times\vec{B})\cdot\vec{C}$$ shows that 
$(\vec{C}\times\vec{B})\cdot\vec{A}=-(\vec{A}\times\vec{B})\cdot\vec{C}$.

It looks also natural for orthogonal vectors $\vec{A}$ and $\vec{B}$ to have
$$|\vec{A}\times\vec{B}|=|\vec{A}|~|\vec{B}|,$$
where $|\vec{A}|^2=\vec{A}\cdot\vec{A}$ is the norm. But then for any two
vectors $\vec{A}$ and $\vec{B}$ the norm $|\vec{A}\times\vec{B}|^2$ 
equals to
$$\left | \left (\vec{A}-\frac{\vec{A}\cdot\vec{B}}
{|\vec{B}|^2}~\vec{B}\right)\times\vec{B}\right |^2=
\left | \vec{A}-\frac{\vec{A}\cdot\vec{B}}{|\vec{B}|^2}~\vec{B}
\right |^2|\vec{B}|^2=|\vec{A}|^2 |\vec{B}|^2-(\vec{A}\cdot\vec{B})^2.$$
Therefore for any two vectors we should have
$$(\vec{A}\times\vec{B})\cdot (\vec{A}\times\vec{B})=(\vec{A}\cdot\vec{A})
(\vec{B}\cdot\vec{B})-(\vec{A}\cdot\vec{B})^2.$$
Now consider
$$|\vec{A}\times (\vec{B}\times\vec{A})-(\vec{A}\cdot\vec{A})\vec{B}+
(\vec{A}\cdot\vec{B})\vec{A}|^2=$$ 
$$=|\vec{A}\times (\vec{B}\times\vec{A})|^2+
|\vec{A}|^4|\vec{B}|^2-(\vec{A}\cdot\vec{B})^2|\vec{A}|^2-
2|\vec{A}|^2(\vec{A}\times (\vec{B}\times\vec{A}))\cdot\vec{B}.$$
But this is zero because
$$|\vec{A}\times (\vec{B}\times\vec{A})|^2=|\vec{A}|^2|\vec{B}\times\vec{A}|^2
=|\vec{A}|^4|\vec{B}|^2-(\vec{A}\cdot\vec{B})^2|\vec{A}|^2$$
and
$$(\vec{A}\times (\vec{B}\times\vec{A}))\cdot\vec{B}=(\vec{B}\times\vec{A})
\cdot (\vec{B}\times\vec{A})=|\vec{A}|^2|\vec{B}|^2-(\vec{A}\cdot\vec{B})^2.$$ 
Therefore we have proven the identity
\begin{equation}
\vec{A}\times (\vec{B}\times\vec{A})=(\vec{A}\cdot\vec{A})\vec{B}-
(\vec{A}\cdot\vec{B})\vec{A}\;.
\label{eq8} \end{equation}
Note that the arrangement of the brackets in the l.f.s. is irrelevant because
vector product is anti-commutative.

However, the familiar identity
\begin{equation}
\vec{A}\times (\vec{B}\times\vec{C})=\vec{B}(\vec{A}\cdot\vec{C})-
\vec{C}(\vec{A}\cdot\vec{B})
\label{eq9} \end{equation}
does not follow in general from the above given intuitively evident properties
of the vector product \cite{18}. To show this, let us introduce a ternary 
product \cite{19} (which is zero if (9) is valid)
$$\{\vec{A},\vec{B},\vec{C}\}=\vec{A}\times (\vec{B}\times\vec{C})-
\vec{B}(\vec{A}\cdot\vec{C})+\vec{C}(\vec{A}\cdot\vec{B}).$$
Equation (8) implies that this ternary product is alternating in its 
arguments. For example
$$0=\{\vec{A}+\vec{B},\vec{A}+\vec{B},\vec{C}\}=\{\vec{A},\vec{B},\vec{C}\}+
\{\vec{B},\vec{A},\vec{C}\}.$$
If $\vec{e}_i,\; i=1\div n$ is some orthonormal basis in the vector space, 
then
$$(\vec{e}_i\times\vec{A})\cdot(\vec{e}_i\times\vec{B})=
((\vec{e}_i\times\vec{B})\times\vec{e}_i)\cdot\vec{A}=
[\vec{B}-(\vec{B}\cdot\vec{e}_i)\vec{e}_i]\cdot\vec{A}$$
and therefore
\begin{equation}
\sum\limits_{i=1}^n(\vec{e}_i\times\vec{A})\cdot(\vec{e}_i\times\vec{B})=
(n-1)\vec{A}\cdot\vec{B}.
\label{eq10} \end{equation}
Using this identity we can calculate
$$\sum\limits_{i=1}^n\{\vec{e}_i,\vec{A},\vec{B}\}\cdot 
\{\vec{e}_i,\vec{C},\vec{D}\}=$$ 
\begin{equation}
=(n-5)(\vec{A}\times \vec{B})\cdot
(\vec{C}\times \vec{D})+2(\vec{A}\cdot\vec{C})(\vec{B}\cdot\vec{D})-
2(\vec{A}\cdot\vec{D})(\vec{B}\cdot\vec{C}).
\label{eq11} \end{equation}
Hence
\begin{equation}
\sum\limits_{i,j=1}^n\{\vec{e}_i,\vec{e}_j,\vec{A}\}\cdot
\{\vec{e}_i,\vec{e}_j,\vec{B}\}=(n-1)(n-3)\vec{A}\cdot\vec{B}
\label{eq12} \end{equation}
and \cite{19}
\begin{equation}
\sum\limits_{i,j,k=1}^n\{\vec{e}_i,\vec{e}_j,\vec{e}_k\}\cdot
\{\vec{e}_i,\vec{e}_j,\vec{e}_k\}=n(n-1)(n-3).
\label{eq13} \end{equation}
The last equation shows that some of $\{\vec{e}_i,\vec{e}_j,\vec{e}_k\}$ 
is not zero if $n>3$. So equation (9) is valid only for the usual 
three-dimensional vector product ($n=1$ case is, of course, not interesting 
because it corresponds to the identically zero vector product). Surprisingly,
we have not very much choice for $n$ even if the validity of (9) is not 
insisted. In fact the dimension of space $n$ should satisfy \cite{18}
(see also \cite{Nieto})
\begin{equation}
n(n-1)(n-3)(n-7)=0.
\label{eq14}\end{equation} 
To prove this statement, let us note that by using
$$\vec{A}\times (\vec{B}\times \vec{C})+(\vec{A}\times \vec{B})\times \vec{C}
=(\vec{A}+\vec{C})\times\vec{B}\times(\vec{A}+\vec{C})-
\vec{A}\times\vec{B}\times\vec{A}-\vec{C}\times\vec{B}\times\vec{C}=$$
$$=2\vec{A}\cdot\vec{C}~\vec{B}-\vec{A}\cdot\vec{B}~\vec{C}-
\vec{B}\cdot\vec{C}~\vec{A}$$
and
$$\vec{A}\times(\vec{B}\times(\vec{C}\times\vec{D}))=
\left . \frac{1}{2} \right [
\vec{A}\times (\vec{B}\times(\vec{C}\times\vec{D}))+
(\vec{A}\times \vec{B})\times(\vec{C}\times\vec{D})-$$
$$-(\vec{A}\times \vec{B})\times(\vec{C}\times\vec{D})-
((\vec{A}\times \vec{B})\times \vec{C})\times\vec{D}+
((\vec{A}\times \vec{B})\times\vec{C})\times\vec{D}+$$
$$+(\vec{A}\times (\vec{B}\times\vec{C}))\times\vec{D}-
(\vec{A}\times (\vec{B}\times\vec{C}))\times\vec{D}-
\vec{A}\times ((\vec{B}\times \vec{C})\times\vec{D})+ $$ 
$$+\vec{A}\times ((\vec{B}\times \vec{C})\times\vec{D})+
\vec{A}\times (\vec{B}\times ( \vec{C}\times\vec{D})) \left . \frac{}{}
\right ]$$
we can check the equality
$$\vec{A}\times \{\vec{B},\vec{C},\vec{D}\}=-\{\vec{A},\vec{B},\vec{C}
\times \vec{D}\}+\vec{A}\times(\vec{B}\times(\vec{C}\times\vec{D}))-
\{\vec{A},\vec{C},\vec{D}\times \vec{B} \}+ $$ $$+
\vec{A}\times(\vec{C}\times(\vec{D}\times\vec{B}))-
\{\vec{A},\vec{D},\vec{B}\times \vec{C} \}+
\vec{A}\times(\vec{D}\times(\vec{B}\times\vec{C}))=$$
$$=-\{\vec{A},\vec{B},\vec{C}\times \vec{D} \}
-\{\vec{A},\vec{C},\vec{D}\times \vec{B} \}
-\{\vec{A},\vec{D},\vec{B}\times \vec{C} \}+
3\vec{A}\times\{\vec{B},\vec{C},\vec{D}\}.$$
The last step follows from
$$3\{\vec{B},\vec{C},\vec{D}\}=\{\vec{B},\vec{C},\vec{D}\}+
\{\vec{C},\vec{D},\vec{B}\}+\{\vec{D},\vec{B},\vec{C}\} =$$ $$=
\vec{B}\times(\vec{C}\times\vec{D})+
\vec{C}\times(\vec{D}\times\vec{B})+
\vec{D}\times(\vec{B}\times\vec{C}).$$
Therefore the ternary product satisfies an interesting identity
\begin{equation}
2~\vec{A}\times \{\vec{B},\vec{C},\vec{D}\}=\{\vec{A},\vec{B},\vec{C}
\times \vec{D}\}+\{\vec{A},\vec{C},\vec{D}\times \vec{B} \}+
\{\vec{A},\vec{D},\vec{B}\times \vec{C} \}
\label{eq15}\end{equation}
Hence we should have
$$4\sum\limits_{i,j,k,l=1}^n|\vec{e}_i\times \{\vec{e}_j,\vec{e}_k,
\vec{e}_l\}|^2=$$ $$=
\sum\limits_{i,j,k,l=1}^n|\{\vec{e}_i,\vec{e}_j,
\vec{e}_k\times \vec{e}_l\}+\{\vec{e}_i,\vec{e}_k,\vec{e}_l\times
\vec{e}_j\}+\{\vec{e}_i,\vec{e}_l,\vec{e}_j\times\vec{e}_k\}|^2.$$
L.h.s. is easily calculated by means of (10) and (13):
$$4\sum\limits_{i,j,k,l=1}^n|\vec{e}_i\times \{\vec{e}_j,\vec{e}_k,
\vec{e}_l\}|^2=4n(n-1)^2(n-3).$$
To calculate the r.h.s. the following identity is useful
\begin{equation}
\sum\limits_{i,j=1}^n\{\vec{e}_i,\vec{e}_j,\vec{A}\}\cdot
\{\vec{e}_i,\vec{e}_j\times\vec{B},\vec{C}\}=-(n-3)(n-6)\vec{A}\cdot
(\vec{B}\times\vec{C})
\label{eq16}\end{equation}
which follows from (11) and from the identity 
$$\sum\limits_{i=1}^n(\vec{e}_i\times\vec{A})\cdot((\vec{e}_i\times\vec{B})
\times\vec{C})=$$ $$=\sum\limits_{i=1}^n(\vec{e}_i\times\vec{A})\cdot
[2\vec{e}_i\cdot\vec{C}~\vec{B}-\vec{B}\cdot\vec{C}~\vec{e}_i-
\vec{e}_i\cdot\vec{B}~\vec{C}-\vec{e}_i\times(\vec{B}\times\vec{C})]=$$
$$=-(n-4)\vec{A}\cdot(\vec{B}\times\vec{C}).$$
Now, having at hand (12) and (16), it becomes an easy task to calculate
$$\sum\limits_{i,j,k,l=1}^n|\{\vec{e}_i,\vec{e}_j,
\vec{e}_k\times \vec{e}_l\}+\{\vec{e}_i,\vec{e}_k,\vec{e}_l\times
\vec{e}_j\}+\{\vec{e}_i,\vec{e}_l,\vec{e}_j\times\vec{e}_k\}|^2=$$
$$=3n(n-1)^2(n-3)+6n(n-1)(n-3)(n-6)=3n(n-1)(n-3)(3n-13).$$
Therefore we should have
$$4n(n-1)^2(n-3)=3n(n-1)(n-3)(3n-13).$$
But 
$$3n(n-1)(n-3)(3n-13)-4n(n-1)^2(n-3)=5n(n-1)(n-3)(n-7)$$
and hence (14) follows.

As we see, the space dimension must equal to the magic number seven \cite{20}
the unique generalization of the ordinary three-dimensional vector product
to become possible.

\section{Seven dimensional Maxwell equations}
Let us assume now that the Maxwell equations (7) are given in a seven 
dimensional space. But one needs a detailed realization of the 
seven-dimen\-sio\-nal vector product to proceed further -- so far we only had
shown that such vector product can exist in principle. For this purpose it
is useful to realize that the vector products are closely related to 
composition algebras \cite{21} (in fact these two notions are equivalent 
\cite{18}). Namely, for any composition algebra with unit element $e$
we can define the vector product in the subspace orthogonal to $e$ by
$x\times y=\frac{1}{2}(xy-yx)$ -- so from standpoint of composition algebra
vector product is just the cummutator.
%divided by 2. 
According to the Hurwitz theorem \cite{22} the only composition
algebras are real numbers, complex numbers, quaternions and octonions. The
first two of them give identically zero vector products. Quaternions produce
the usual three-dimensional vector product. The seven-dimensional vector 
product is generated by octonions  \cite{23}. It is interesting to note that
this seven-dimensional vector product is covariant with respect to the 
smallest exceptional Lie group $G_2$ \cite{24} which is the automorphism
group of octonions.

Using the octonion multiplication table \cite{23} one can concretize the
seven-dimensional vector product as follows
\begin{equation}
\vec{e}_i\times\vec{e}_j=f_{ijk}\vec{e}_k\equiv
\sum\limits_{k=1}^7 f_{ijk}\vec{e}_k,\;\;\;
i,j=1\div 7.
\label{eq17}\end{equation}
Where $f_{ijk}$ is completely antisymmetric $G_2$-invariant tensor and the
only nonzero independent components are
$$f_{123}=f_{246}=f_{435}=f_{651}=f_{572}=f_{714}=f_{367}=1.$$
Therefore, for example, the second equation in (7) is equivalent to the system
$$-\frac{\partial B_1}{\partial t}=\frac{\partial E_3}{\partial x_2}-
\frac{\partial E_2}{\partial x_3}+ \frac{\partial E_5}{\partial x_6}-
\frac{\partial E_6}{\partial x_5}+ \frac{\partial E_7}{\partial x_4}-
\frac{\partial E_4}{\partial x_7},$$
$$-\frac{\partial B_2}{\partial t}=\frac{\partial E_1}{\partial x_3}-
\frac{\partial E_3}{\partial x_1}+ \frac{\partial E_6}{\partial x_4}-
\frac{\partial E_4}{\partial x_6}+ \frac{\partial E_7}{\partial x_5}-
\frac{\partial E_5}{\partial x_7},$$
$$-\frac{\partial B_3}{\partial t}=\frac{\partial E_2}{\partial x_1}-
\frac{\partial E_1}{\partial x_2}+ \frac{\partial E_4}{\partial x_5}-
\frac{\partial E_5}{\partial x_4}+ \frac{\partial E_7}{\partial x_6}-
\frac{\partial E_6}{\partial x_7},$$
$$-\frac{\partial B_4}{\partial t}=\frac{\partial E_2}{\partial x_6}-
\frac{\partial E_6}{\partial x_2}+ \frac{\partial E_5}{\partial x_3}-
\frac{\partial E_3}{\partial x_5}+ \frac{\partial E_1}{\partial x_7}-
\frac{\partial E_7}{\partial x_1},$$
$$-\frac{\partial B_5}{\partial t}=\frac{\partial E_3}{\partial x_4}-
\frac{\partial E_4}{\partial x_3}+ \frac{\partial E_6}{\partial x_1}-
\frac{\partial E_1}{\partial x_6}+ \frac{\partial E_2}{\partial x_7}-
\frac{\partial E_7}{\partial x_2},$$
$$-\frac{\partial B_6}{\partial t}=\frac{\partial E_4}{\partial x_2}-
\frac{\partial E_2}{\partial x_4}+ \frac{\partial E_1}{\partial x_5}-
\frac{\partial E_5}{\partial x_1}+ \frac{\partial E_3}{\partial x_7}-
\frac{\partial E_7}{\partial x_3},$$
$$-\frac{\partial B_7}{\partial t}=\frac{\partial E_5}{\partial x_2}-
\frac{\partial E_2}{\partial x_5}+ \frac{\partial E_4}{\partial x_1}-
\frac{\partial E_1}{\partial x_4}+ \frac{\partial E_6}{\partial x_3}-
\frac{\partial E_3}{\partial x_6}.$$

Now it is time to realize that there are some subtleties in Feynman's
derivation of these seven dimensional Maxwell equations because
$f_{ijk}f_{kmn}\ne\delta_{im}\delta_{jn}-\delta_{in}\delta_{jm}.$ Instead
we have
\begin{equation}
f_{ijk}f_{kmn}=g_{ijmn}+\delta_{im}\delta_{jn}-\delta_{in}\delta_{jm}
\label{eq18} \end{equation}
where
$$g_{ijmn}=\vec{e}_i\cdot\{\vec{e}_j,\vec{e}_m,\vec{e}_n\}.$$
In fact $g_{ijmn}$ is completely antisymmetric $G_2$-invariant tensor.
For example
$$g_{ijmn}=\vec{e}_i\cdot\{\vec{e}_j,\vec{e}_m,\vec{e}_n\}=
-\vec{e}_i\cdot\{\vec{e}_m,\vec{e}_j,\vec{e}_n\}=$$ $$=
-\vec{e}_i\cdot(\vec{e}_m\times(\vec{e}_j\times\vec{e}_n))+(\vec{e}_i\cdot
\vec{e}_j)(\vec{e}_m\cdot\vec{e}_n)-(\vec{e}_i\cdot\vec{e}_n)
(\vec{e}_m\cdot\vec{e}_j)=$$ $$=-\vec{e}_m\cdot(\vec{e}_i\times
(\vec{e}_n\times\vec{e}_j))+(\vec{e}_i\cdot\vec{e}_j)(\vec{e}_m
\cdot\vec{e}_n)-(\vec{e}_i\cdot\vec{e}_n)(\vec{e}_m\cdot\vec{e}_j)=$$ $$=
-\vec{e}_m\cdot\{\vec{e}_i,\vec{e}_n,\vec{e}_j\}=
-\vec{e}_m\cdot\{\vec{e}_j,\vec{e}_i,\vec{e}_n\}=-g_{mjin}.$$
The only nonzero independent components are
$$g_{1254}=g_{1267}=g_{1364}=g_{1375}=g_{2347}=g_{2365}=g_{4576}=1.$$

Let us modify the Feynman's derivation in such a way that it will be valid
in seven-dimensional case also. For the magnetic field we retain the 
same definition:
\begin{equation}
B_i=-\frac{im^2}{2\hbar}f_{ijk}[v_j,v_k]=\frac{1}{2}f_{ijk}F_{jk}
\label{eq19} \end{equation}
where we had introduced stress tensor
$$F_{ij}(x,t)=-\frac{im^2}{\hbar}[v_i,v_j].$$
But now the stress tensor is not completely determined by the magnetic field.
Instead we have from (18)
\begin{equation}
F_{ij}=f_{ijk}B_k-\frac{1}{2}g_{ijkl}F_{kl}.
\label{eq20}\end{equation}
This leads to a modification of the Lorentz force law. Indeed, integration of
(4) gives
$$F_i(x,v,t)=E_i(x,t)+<F_{ij}(x,t)v_j>$$
and then (20) implies
\begin{equation}
F_i(x,v,t)=E_i(x,t)+f_{ijk}<v_jB_k>-~\frac{1}{2}g_{ijkl}<v_jF_{kl}>.
\label{eq21}\end{equation}
So we have an extra term in the Lorentz force law! Note that (21) can be
rewritten as
$$\vec{F}(x,v,t)=\vec{E}(x,t)+<\vec{v}\times\vec{B}(x,t)>+~
\frac{im^2}{\hbar}\{\vec{v},\vec{v},\vec{v}\}.$$

A variant of the proof of the Faraday induction law which remains valid in 
the seven-dimensional case goes like this. Using 
$F_i=E_i+<F_{ij}v_j>$ we can rewrite the
equation
$$\frac{\partial \vec{B}}{\partial t}+<\vec{v}\cdot\nabla\vec{B}>=
-\frac{im}{\hbar}(\vec{F}\times\vec{v}+\vec{v}\times \vec{F})$$
in the following way
$$\frac{\partial B_i}{\partial t}+<\vec{v}\cdot\nabla\vec{B}>_i=
-({\mathrm rot}\vec{E})_i-\frac{im}{\hbar}f_{ijk}[v_j,<F_{km}v_m>].$$
But
$$f_{ijk}[v_j,<F_{km}v_m>]=$$ $$=\left .\frac{1}{2}f_{ijk}
\right ( [v_j,F_{km}]v_m+
v_m[v_j,F_{km}]+F_{km}[v_j,v_m]+[v_j,v_m]F_{km}\left )\frac{}{}\right ..$$
The last two terms give zero contribution, because
$$f_{ijk} (F_{km}[v_j,v_m]+[v_j,v_m]F_{km})\sim f_{ijk} (F_{km}F_{jm}+
F_{jm}F_{km})=0.$$
Using the Jacoby identity we can transform
$$f_{ijk}[v_j,F_{km}]=-\frac{im^2}{\hbar}f_{ijk}[v_j,[v_k,v_m]]=
\frac{im^2}{\hbar}f_{ijk}([v_k,[v_m,v_j]]+ $$ $$+[v_m,[v_j,v_k]])=
f_{ijk}([v_k,F_{jm}]+[v_m,F_{kj}])=-f_{ijk}[v_j,F_{km}]-2[v_m,B_i].$$
Therefore
$$f_{ijk}[v_j,F_{km}]=-[v_m,B_i]=\frac{i\hbar}{m}~\frac{\partial B_i}
{\partial x_m}$$
and
$$-\frac{im}{\hbar}f_{ijk}[v_j,<F_{km}v_m>]=<\vec{v}\cdot\nabla\vec{B}>_i.$$
Hence the desired Faraday law follows.
\section{Concluding remarks}  
So Feynman's derivation of Maxwell equations really leads to the unique new
theory. We used a generalization of the vector product to come to this 
conclusion. This generalization is only possible in a seven-dimensional
space and is closely related to octonions -- the largest composition algebra
which ties together many exceptional structures in mathematics \cite{24}.
In more general case of p-fold vector products some more possibilities
arise \cite{21,25,26} but without any clear connection with Maxwell equations
or Feynman's derivation. 

It is amusing that the Maxwell equations, which harbour many beautiful
mathematical concepts \cite{27}, have connections with octonions also. This
connection suggests that there might be something exceptional in the
seven-dimensional electrodynamics. Note that the extra term we obtained in 
the Lorentz force law has its roots in the non-associativity of octonion 
algebra.
The important question, however, is whether the seven-dimensional Maxwell 
equations have any contact with reality, or they should be considered just as 
a nice mathematical curio. The best way to express my uncertainty and  
confusion about the answer on this question is to provide the concluding 
fragment from G.~A.~Miller's essay \cite{20}. 

``And finally, what about the magical number seven? What about the seven 
wonders of the world, the seven seas, the seven deadly sins, the seven 
daughters of Atlas in the Pleiades, the seven ages of man, the seven levels 
of hell, the seven primary colors, the seven notes of the musical scale, 
and the seven days of the week? What about the seven-point rating
scale, the seven categories for absolute judgment, the seven objects in the 
span of attention, and the seven digits in the span of immediate memory? For 
the present I propose to withhold judgment. Perhaps there is something deep 
and profound behind all these sevens, something just calling out for us to 
discover it. But I suspect that it is only a pernicious, Pythagorean 
coincidence''. 

\section*{Acknowledgements}
This paper was inspired by  R.~S.~Garibaldi's lectures \cite{28}.
Support from the INTAS grant No. 00-00679 is also acknowledged.
%\newpage

\end{document}